\documentstyle[preprint,revtex,eqsecnum]{aps}
\begin{document}
\draft
\preprint{Alberta-Thy-36-92}
\begin{title}
QUANTUM EFFECTS IN BLACK HOLE INTERIORS
\end{title}
\author{Warren G. Anderson, Patrick R. Brady, Werner Israel,  \\and Sharon
M. Morsink}%
\begin{instit}
Canadian Institute for Advanced Research Cosmology Program, \\
Theoretical Physics Institute,
 University of  Alberta,
Edmonton, Alberta, Canada T6G 2J1
\end{instit}
%
%
\begin{abstract}
The Weyl curvature inside a black hole formed in a generic collapse grows,
classically without bound, near to the inner horizon, due to partial
absorption and blueshifting of the radiative tail of the collapse.  Using a
spherical model, we examine how this growth is modified by quantum effects
of conformally coupled massless fields.
\end{abstract}
\pacs{PACS numbers: 9760L, 0460}

\narrowtext

%
%
Classical models of generic black hole interiors \cite{ref:1}-\cite{ref:8}
have made progress in unravelling the nature of the internal geometry up to
the onset of singular behavior at the inner (Cauchy) horizon.  At this
lightlike hypersurface, which corresponds to infinite external advanced
time, the {``Coulomb component''} $|\Psi_2|$ of the Weyl curvature diverges
exponentially with advanced time.  (For spherical symmetry, $|\Psi_2|\approx
m/r^3$ in terms of the Schwarzschild local mass function $m$.)

Here we report on a first attempt to gauge the influence of quantum effects
on this scenario; in particular, to examine whether vacuum polarization and
pair creation will act so as to damp the classical rise of curvature and
possibly limit it to sub-Planck values.

The classical divergence of $|\Psi_2|$ is caused by the partial absorption,
and infinite blueshifting~\cite{ref:9} at the Cauchy horizon~(CH), of
gravitational radiation from the radiative tail~\cite{ref:10} of the
collapse,
in concert with radiative outflow from the star as it shrinks within the
hole. (The outflow has the merely catalytic role of focussing the generators
and initiating the contraction of the CH.)

Ori~\cite{ref:3} has devised a simple model, involving a spherical charged
black hole, which appears to capture the essence of the physics.  The influx
of gravitational waves is modelled by a radial stream of lightlike
particles.  (This ``graviton'', ``effective stress-energy''
description~\cite{ref:11} is justified by the large blueshift.)  The outflow
is treated schematically as a thin, transparent lightlike shell $\Sigma$
within and parallel to the event horizon~(Fig. 1).

The metric in each of the domains ${\cal V}_-$ and ${\cal V}_+$ separated by
$\Sigma$ then has the charged Vaidya form for pure inflow:
		\begin{eqnarray}
		(ds^2)_{\pm} &=& dv_{\pm}(2\, dr - f_{\pm} \, dv_{\pm}) +
r^2\,
		d\Omega^2\: ,\mbox{\hspace{.25in}}\nonumber\\
		f_{\pm} &=& 1 -
		\frac{2m_{\pm}}{r} +
		\frac{e^2}{r^2}
		\label{eqn:1}\: .
		\end{eqnarray}
The advanced time parameters $v_+$ and $v_-$ are unequal.  They are related
by noting that the equations of $\Sigma$ with respect to the two abutting
coordinate systems are
		\begin{equation}
		f_+\, dv_+ = f_-\, dv_- = 2\, dr \mbox{\hspace{.25in}}
		{\rm along} \ \Sigma\:.
		\label{eqn:2}
		\end{equation}
Continuity of the influx across $\Sigma$ requires that
		\begin{equation}
		f_+^{-2} \frac{dm_+}{dv_+} =   f_-^{-2} \frac{dm_-}{dv_-}\:
{}.
		\label{eqn:3}
		\end{equation}
{}From (\ref{eqn:2}) and (\ref{eqn:3}),
		\begin{equation}
		\frac{dm_+}{f_+} = \frac{dm_-}{f_-}\mbox{\hspace{.25in}}
		{\rm along} \ \Sigma \: ,
		\end{equation}
which clearly shows the divergence of the interior mass function $m_+$ as
one approaches the CH~($f_-\rightarrow 0$, $v_-\rightarrow\infty$).

	The ansatz
		\begin{equation}
		m_-(v_-) = m_0 - ({\rm const.})\times v_-^{-(p-1)}
		\end{equation}
reproduces the correct, power-law decay $dm_-/dv_- \sim v_-^{-p}$ of the
externally observed gravitational wave flux (p=12 for quadrupole
waves~\cite{ref:10}).

Integration of the foregoing equations is now straightforward, and
yields~\cite{ref:3,ref:8} to leading order (with $m_+$ written simply as $m$
from here on),
		\begin{eqnarray}
		m(v_+) &=& m_0 |\ln\,(-v_+/m_0)|^{-p}(-v_+/m_0)^{-1}
		\mbox{\hspace{.25in}}
		(v_+/m_0
		\rightarrow 0^-)\: , \label{eqn:4}\\
		-v_+ &=& ({\rm const.}) \times \exp (-\kappa_0v_-)
		\mbox{\hspace{.25in}}(v_-
		\rightarrow
		\infty) \: ,
		\end{eqnarray}
Here, $\kappa_0 = (m_0^2 - e^2)^{1/2}/ r_0^2$ is the surface gravity (and
$r_0$ the radius) of the initial, static segment of the inner horizon in
${\cal V}_-$, and we have set a dimensionless numerical coefficient in
(\ref{eqn:4}) (depending on the luminosity and initial deformation of the
collapsing star) equal to unity.

A Vaidya geometry with metric of the form (\ref{eqn:1}) has the Ricci
curvature
		\begin{equation}
		R_{\alpha\beta} \approx 2\dot{m}(v_+)r^{-2}
		(\partial_{\alpha}v_+ )
		(\partial_{\beta}v_+ ) \label{eqn:5}
		\end{equation}
where the contribution of the electrostatic field has been ignored, and
		\begin{equation}
		- \Psi_2 = \frac{1}{2}C{^{\theta\phi}}_{\theta\phi} =
		[m(v_+ ) - e^2/r] r^{-3} \label{eqn:6}
		\end{equation}
is the sole non-vanishing Newman-Penrose component of the Weyl curvature.

	To study effects of quantum corrections we aim to estimate the
expectation value $\left<T_{\alpha\beta}\right>$ (in the Unruh state) of the
stress-energy for conformally coupled massless fields on this Vaidya
background when the mass function has classically the diverging
form~(\ref{eqn:4}).

	In general, finding $\left<T_{\alpha\beta}\right>$ is a problem of
notorious difficulty~\cite{ref:12}.  It becomes  tractable in the present
instance because of a number of special circumstances: (i) We are only
interested in the asymptotic form of $\left<T_{\alpha\beta}\right>$ near the
CH. (ii) The singularity is relatively mild ($\Psi_2$ is a diverging but
integrable function of $v$)~\cite{ref:3,ref:4}. (iii) The lightlike
character of (\ref{eqn:5}) means that ordinarily dominant terms nonlinear in
$R_{\alpha\beta}$ actually vanish. (iv)~The special form (\ref{eqn:4}) of
the mass function means that, of two terms containing the same total number
of $m$-factors and $v$-derivatives, the term with the smaller number of
$m$-factors is dominant; e.g. $\ddot{m} >> m\dot{m}/r^2$.  (The ratio of the
two sides is ``merely'' a logarithmic factor; however this factor does
become infinite as $v \rightarrow 0^-$, and for $v=-10^{10}$ Planck times in
a solar mass black hole, it has already grown to $10^{23}$~!)

	This conjunction of circumstances permits us to treat the geometry
as a linear perturbation of flat space.  The terms linear in derivatives of
$m(v)$, which we retain, actually dominate the neglected nonlinear terms.

	Barvinsky and Vilkovisky~\cite{ref:13} have developed the first
stages of a systematic general formalism applicable in such circumstances,
i.e., terms linear (and, more generally of algebraically lower order) in
curvature and derivatives dominate over higher order terms having the same
dimension.  In the work reported here, we shall apply a simpler, first-order
prescription due to Horowitz~\cite{ref:14}.

	First, we address and remove a potential source of difficulty.  The
Weyl curvature approaches Planck levels, $|\Psi_2|\rightarrow t_p^{-2}$, as
the advanced time coordinate $v_+\rightarrow v_p <0$, where
		\begin{equation}
		|v_p| = b^{-2} \epsilon t_p [\ln\, (b/\epsilon^2)]^{-p}\; ,
		\ \ \ b\equiv r_0/m_0\: ,\ \  \epsilon \equiv t_p / r_0 \: ,
		\label{eqn:7}\end{equation}
and $t_p = 10^{-43}$s is the Planck time. One may therefore hope that, for
$|v_+| >> |v_p|$, effects of quantum gravity will remain small, and the
spacetime geometry effectively classical. The strongly divergent expression
(\ref{eqn:5}) raises some doubt on this score, suggesting that Ricci
curvature may dominate Weyl curvature and surpass Planck levels at times
much earlier than $v_p$.  This statement is, of course, not coordinate
independent: an observer with four velocity $u^{\alpha}$ measures a Ricci
curvature $R_{\alpha\beta}u^{\alpha}u^{\beta}$ which can be arbitrarily
small if he falls inward at nearly the speed of light, whereas the Weyl
curvature scalar is boost-invariant.  Nonetheless, it might be cause for
worry that the growth of Ricci curvature in some physically ``reasonable''
frame (however defined) will make our semiclassical analysis meaningless.

	It is easy to allay such doubts for all practical purposes.  The
conformal transformation
		\begin{equation}
		ds^2 = (r/r_0)^2 ds_*^2 \label{eqn:8}
		\end{equation}
generates a new Ricci tensor $R^*_{\alpha\beta}$ free of the strongly
divergent $\dot{m}$ terms (see (\ref{eqn:11}) below), while merely
multiplying $\Psi_2$ by a factor $(r/r_0)^2$ of order unity.  We shall
obtain $\left<T_{\alpha\beta}\right>$ in the conformal metric, then use
Page's formula~\cite{ref:15} to transform back to the physical metric.

	It is helpful to rescale the null coordinate $v_+$ so that it more
nearly represents Planck scales.  We set
		\begin{equation}
		v\equiv\epsilon v_+\: , \ \  u \equiv 2\epsilon r_0^2/r\: , \
		\ \overline{\theta}\equiv \epsilon^{-1}\theta\: ,\ \  x+ i\, y
		\equiv t_p \epsilon^{-1} e^{i\,\phi}\, \sin
\epsilon\overline{\theta}
		\: ,
		\end{equation}
so that $r^2_0d\Omega^2 \approx dx^2 + dy^2$, i.e. the sphere $r=r_0$ is
nearly flat on Planck scales.  Further, since $m(v_+)$ is much larger than
$|e|$ and $r$ near the CH, we may use the approximation $f\approx-2m/r$.

	The conformal metric now takes the Kerr-Schild (flat plus lightlike)
form
		\begin{eqnarray}
		ds_*^2 &=& -dudv + dx^2 + dy^2 + 2L(u,v)dv^2
		\label{eqn:9}\: ,\\
		L  &=& \frac{1}{8} ku^3m(v_+)\: , \ \ k = (t_pr_0^3)^{-1}\:
,
		\label{eqn:10}
		\end{eqnarray}
which is manifestly almost flat for $|\Psi_2| << t_p^{-2}$.  The Ricci
curvature for (\ref{eqn:9}) is
		\begin{equation}
		R_{\alpha\beta}^* = 4L_{uu} (2L\, l_{\alpha}l_{\beta} -
		l_{(\alpha
		}n_{\beta)}) \label{eqn:11}
		\end{equation}
where $l_{\alpha} = -\partial_{\alpha}v$, $n_{\alpha} = -\partial_{\alpha}u
$ and the $u$-subscripts denote partial derivatives.

	A general argument due to Horowitz~\cite{ref:14} shows that, to
linear order in a nearly flat spacetime, the in-in vacuum expectation value
$\left<T_{\mu\nu}\right>$ of the stress-energy for a conformally coupled
massless field is given in terms of the retarded integral over the past
light-cone of the point $x$ in the flat background;
		\begin{equation}
		t_p^{-2}\left<T_{\mu\nu}\right>_{\rm ren} = a\int\!\!
		H(x-x')A_{\mu\nu}(x')d^4x'
		 + \alpha A_{\mu\nu}(x) + \beta
		B_{\mu\nu}(x)
		\label{eqn:12}\: .
		\end{equation}
Here, $a$ is a positive numerical coefficient whose value is known for
different spins~\cite{ref:14}; $\alpha$ and $\beta$ are arbitrary numbers;
$A_{\mu\nu}$ and $ B_{\mu\nu}$ are the (linearized) variational derivatives
$\delta /\delta g^{\mu\nu}$ of the actions associated with
$C_{\alpha\beta\gamma\delta}^2$ and $R^2$ respectively -- explicitly,
		\begin{eqnarray}
		A_{\mu\nu} &=& -2\Box G_{\mu\nu} - \frac{2}{3}
		 G{^{\alpha}}_{\alpha\, ,\, \mu\nu} + \frac{2}{3}
		 \eta_{\mu\nu}\Box
	 	G{^{\alpha}}_{\alpha} \nonumber\\
		 B_{\mu\nu} &=& 2\eta_{\mu\nu}\Box G{^{\alpha}}_{\alpha} - 2
		 G{^{\alpha}}_{\alpha\, ,\, \mu\nu} \: .\label{eqn:13}
	 	\end{eqnarray}
The past light-cone distribution $H$ is given by
		\begin{equation}
		H(x-x') = \delta'(\sigma) \theta (t-t')\: , \ \  \ \sigma
		\equiv
		\frac{1}{2} (x_{\mu}-x'_{\mu})^2\: .
		\end{equation}
Explicitly, for any function $f$,
		\begin{equation}
		\int \!\! H(x-x') f(x') d^4x' = \int_{4\pi}\!\!
		d\Omega\int_{-\infty}^0\!\!
		dU
		\left[ \frac{\partial f}{\partial U}
		\ln\,\left(\frac{-U}{\lambda}\right)
		 + \frac{1}{2}\frac{\partial
		f}{\partial V}
		\right]_{V=0}\: , \label{eqn:14}
		\end{equation}
where $U$, $V$, $d\Omega$ are spherical lightlike coordinates centered on
$x$, so that the past lightcone of $x$ is $V=0$.  The arbitrary length scale
$\lambda$ may be considered to reflect the arbitrariness of $\alpha$ in
(\ref{eqn:12}), and could be adjusted so as to absorb $\alpha$.

	The in-vacuum expression (\ref{eqn:12}) needs to be supplemented by
a local, conserved tensor representing initial conditions appropriate to the
Unruh state for an evaporating black hole.  Inside the hole, this is just
the lightlike influx of negative energy  that accompanies the thermal
outflux to infinity~\cite{ref:12}.  However, this remains negligible up to
the moment $v_+ = v_p$ when the classical curvature becomes Planckian if the
black hole is larger than 100kg~\cite{ref:4}, and it will therefore be
ignored.

	According to (\ref{eqn:13}), (\ref{eqn:11}) and (\ref{eqn:10}), the
functions to be inserted in (\ref{eqn:12}) are

		\begin{eqnarray}
		A_{uu} &=& B_{uu} = 0\:, \ \ A_{vv} = \frac{1}{3} B_{vv} = 4
		kud^2m/dv^2\: , \label{eqn:15} \\
		A_{uv} &=&  \frac{1}{3} B_{uv} = -\frac{1}{2} A_{xx}
		= -\frac{1}{2} A_{yy} = \frac{1}{12} B_{xx} = \frac{1}{12}
		B_{yy}
		=4kdm/dv\: ,\nonumber
		\end{eqnarray}
with $m$ given by (\ref{eqn:4}).

	It is now straightforward to express the spherical lightlike
coordinates $U$, $V$, $d\Omega$ of (\ref{eqn:14}) in terms of the plane
coordinates of (\ref{eqn:9}) and then evaluate the integrals approximately
by Laplace's method.

	From (\ref{eqn:15}) it is immediate that $\left<T_{uu}\right> = 0$.
Transforming back to the physical metric (\ref{eqn:8}) does not change this
result significantly.  Page's formula~\cite{ref:15} yields for the quantum
outflux
		\begin{equation}
		\left<T^{\alpha\beta}\right> (\partial_{\alpha}v_+)
		(\partial_{\beta}v_+)  \sim t_p^{2} m/r^5\: ,
		\end{equation}
which is of order $t_p^2\Psi_2$ times the classical outflux from the
collapsing star. The smallness of this result has the important consequence
that the classical picture of the CH contracting very slowly (on Planck
scales) under irradiation by the star is not affected by the quantum
corrections up to the time when curvatures become
Planckian~(cf~\cite{ref:7}).

	The dominant contribution to $\left< T_{\alpha\beta}\right> $ comes
from the logarithmic term in (\ref{eqn:14}) with $f= A_{vv}$:
		\begin{equation}
		\left<T_{v_+ v_+}\right> \approx a_* t_p^2 r^{-3} \ddot{m}
		(v_+)\ln\,(-v_+/\lambda) \label{eqn:16}
		\end{equation}
where $a_*$ is a positive number, approximately $a_* = \frac{729}{8}
\left(\frac{2\pi^3}{3}\right)^{1/2} a $.  (The result quoted is for the
physical metric.  The only effect of the conformal transformation on the
leading term of $\left<T_{v_+ v_+}\right>$ is to rescale  the arbitrary
length scale $\lambda$.)

	There are now two essentially different possibilities.  If $\lambda
>> |v_p|$ ($v_p < 0$ is the coordinate time (\ref{eqn:7}) at which the
classical curvature approaches Planck levels), the logarithm in
(\ref{eqn:16}) is negative as $v_+$ approaches $v_p$,  and (\ref{eqn:16})
predicts damping of the classical growth of curvature due to quantum
effects.  If, on the other hand, $\lambda
\:\vtop{\vskip-7pt\hbox{$<$}\vskip-8pt \hbox{$\sim$}}\: |v_p|$, quantum
effects would further destabilize the classical plunge toward a curvature
singularity.

		Unfortunately, this ambiguity cannot  be resolved within the
present theoretical framework, which provides no information about
$\lambda$~\cite{ref:16}.  The quantum theory of massless fields propagating
on a fixed classical background has no inherent length-scale.

			Something further can be said if one is willing to
entertain an arguable hypothesis about the origin of this incompleteness of
the semi-classical theory~\cite{ref:16}.

			Quantum effects of the gravitational field itself
have not yet been included.  A successful (renormalizable or finite) quantum
theory of the gravitational field would be expected to have the effect, at
moderate curvatures, of modifying the Einstein-Hilbert Lagrangian by terms
quadratic in curvature,
		\begin{equation}
		16\pi L_G = t_p^{-2} R + \alpha_1
C_{\alpha\beta\gamma\delta}^2
		+
		\beta_1R^2\: ,
		\end{equation}
where $\alpha_1$ and $\beta_1$ are constants of order unity.  (General
arguments due to Fradkin and Vilkovisky~\cite{ref:17} suggest that
$\alpha_1$ is negative.)  The added terms induced in the effective field
equations,
		\begin{equation}
		G_{\mu\nu} + t^2_p(\alpha_1 A_{\mu\nu} + \beta_1B_{\mu\nu} )
=
		8\pi \left\{ T^{\rm class}_{\mu\nu} +
\left<T_{\mu\nu}\right>
		\right\} \label{eqn:17}
		\end{equation}
are of the same form as the terms left arbitrary in (\ref{eqn:12}).  This
suggests that the incompleteness of the semi-classical theory is related to
the neglect of quantum gravitational effects.

	It is arguable that a quantum theory of massless fields which
includes gravity would become a {\em complete} theory, with the net
coefficients of $A_{\mu\nu}$ and $B_{\mu\nu}$ determined~\cite{ref:16}.

		Suppose this is true.  Adjust $\lambda$ in (\ref{eqn:14}) so
that $\alpha =0$ in (\ref{eqn:12}).  Then $\alpha_1$ in (\ref{eqn:17}) is
expected to be of order unity, and the term $\alpha_1 A_{\mu\nu}$ may be
interpreted as representing effects associated with gravitational vacuum
polarization.  Now, it is reasonable to expect that, once the quantum
gravitational degrees of freedom are activated (i.e. when $v_+\approx v_p$),
gravitons will have effects not too disimilar from photons and other
massless fields, i.e. that
		\begin{equation}
		\left<T_{\mu\nu}\right>\sim -\alpha_1 t^2_p A_{\mu\nu} \  \
\
		{\rm for} \ \ v_+\approx v_p\:.
		\end{equation}
Comparison of (\ref{eqn:15}) and (\ref{eqn:16}) now shows that the
logarithmic factor is of order unity, i.e. that $\lambda\approx |v_p|$. (The
circumstance that this is much shorter than a Planck time $t_p$ has no
physical significance, because the scale of the null coordinate $v_+$ (which
$\lambda$ normalizes) has no intrinsic local meaning.)

	If this conclusion is correct, (\ref{eqn:16}) should be interpreted
as an intensification (rather than a damping) of the classical influences
tending to produce a curvature singularity at the CH, at least for the Ori
spherical model considered here.  It would indicate strongly that the
ultimate, quantum stage of evolution of the hole is inaccessible to
semi-classical considerations. A detailed account of this work is in
preparation.

	It is a pleasure to thank R. Balbinot, C. Barrab\`es, A. Barvinsky,
R. Camporesi, S. Droz, D. Page and E. Poisson for stimulating discussions.
This work was supported by the Canadian Institute for Advanced Research and
the Natural Sciences and Engineering  Research Council of Canada.  One of us
(SMM) wishes to acknowledge financial support from the Alberta Heritage
fund, an Avadh Bhatia Fellowship and the Killam Foundation.

\figure{
Ori model.  Infalling radiation passes through a transparent, ``outgoing''
lightlike shell $\Sigma$ inside a charged spherical hole.  EH is the event
horizon, and CH the Cauchy horizon.}

	\end{document}